\documentclass[useAMS,usenatbib,twocolumn]{mn2e}
\usepackage{graphicx}
\usepackage{amssymb}
\usepackage{amsmath}
\usepackage{fleqn}
\usepackage{subfigure}
\usepackage{float}

\def\half{{\frac{1}{2}}}
\def\threeh{{\frac{3}{2}}}

\def\ga{{g_{\rm asy}}}

\def\upartial{\partial}
\def\rmin{{r_{-}}}

\def\aj{AJ}
\def\mnras{MNRAS}
\def\apj{ApJ}
\def\apjl{ApJL}
\def\nat{Nat}
\def\na{New Ast}
\def\aap{AA}

\title[Is there substructure around M87?]
        {Is there substructure around M87?}
\author[Oldham \& Evans]
   {L.~J.~Oldham$^1$\thanks{Email: loldham@ast.cam.ac.uk,nwe@ast.cam.ac.uk} \& N.~W.~Evans$^1$
  \medskip
  \\$^1$Institute of Astronomy, University of Cambridge, Madingley Road,
       Cambridge, CB3 0HA, UK}
\begin{document}

\date{Accepted  Received ; in original form }

\pagerange{\pageref{firstpage}--\pageref{lastpage}} \pubyear{2015}

\maketitle

\label{firstpage}

\begin{abstract}
We present a general method to identify infalling substructure in
discrete datasets with position and line-of-sight velocity data. We
exploit the fact that galaxies falling onto a brightest cluster galaxy
(BCG) in a virialised cluster, or dwarf satellites falling onto a
central galaxy like the Milky Way, follow nearly radial orbits. If the
orbits are exactly radial, we show how to find the probability
distribution for a satellite's energy, given a tracer density for the
satellite population, by solving an Abel integral equation.  This is
an extension of \citet{Ed16}'s classical formula for the isotropic
distribution function.  When applied to a system of galaxies,
clustering in energy space can then be quantified using the
Kullback-Leibler divergence, and groups of objects can be identified
which, though separated in the sky, may be falling in on the same
orbit. This method is tested using mock data and applied to the
satellite galaxy population around M87, the BCG in Virgo, and a number
of associations are found which may represent infalling galaxy groups.
\end{abstract}


\section{Introduction}\label{sec:introduction}

In the hierarchical model of galaxy formation, elliptical galaxies and
the stellar haloes of spiral galaxies are built up gradually by
prolonged periods of accretion. In this picture, early-type galaxies
form during an early phase of dissipational star formation fed
by gas-rich mergers, creating compact cores, while their subsequent
evolution is mainly driven by abundant minor mergers, feeding the
outer regions of the galaxy and bringing them onto the local size-mass
relation \citep[e.g.][]{Shen2003, Na09, Hopkins2010}.

This phenomenon is even more important in clusters, where the
brightest cluster galaxies (BCGs) that reside in their centres are
thought to have acquired the majority of their stars through the
accumulation and subsequent destruction of satellite
galaxies~\citep{La13,Co15}.  BCGs have haloes that are embedded in the
intracluster light, and it is not obvious whether a useful distinction
between the halo and the intra-cluster medium is even
possible~\citep{Go05}; the disruption of satellite galaxies as they
are funnelled by dynamical friction from the intra-cluster medium onto
the central BCG then creates an extensive envelope. Indeed, it is
possible that the progenitors of today's BCGs are the extremely
compact `red nuggets' that have been observed at redshifts $z \sim 2$
\citep[e.g.][]{Trujillo2007, vanDokkum2008}. The evolution of the BCG
is therefore dominated by the addition of stars and globular clusters
that have formed outside the BCG itself.

In the Local Group, a much lower-mass example compared to galaxy
clusters, the biggest members show convincing evidence of a structured
satellite galaxy population. For instance, the Large Magellanic Cloud,
together with its satellites, may be part of an extended
group that is on its first infall onto the Milky Way~\citep{Ko15},
while M31 seemingly has an extended thin disk of
satellites~\citep{Ib13} that may be the result of group infall and
accretion~\citep[e.g.][]{Bo14}. In these systems, detailed searches
for substructure are made possible by the high-quality six-dimensional
phase-space data that are available for individual stars
\citep[e.g.][]{Xue2011}.

Given these findings, the extreme densities found in cluster
environments make it very likely that the outer parts of BCGs are also
permeated with radially infalling satellite groups. However,
substructure identification in these much more distant systems cannot
be carried out using the same methods as in the Local Group as it is
not possible to resolve individual stars, and bright stellar proxies
such as globular clusters, planetary nebulae and satellite galaxies
must be used instead. Another difficulty is that these studies must be
carried out using projected data, as only three of the full
six-dimensional phase space coordinates are usually available --
namely, position on the sky and line-of-sight velocity -- and this
gives rise to much larger uncertainty. Nevertheless, a number of recent
studies have used globular cluster kinematics to find evidence for
recent accretion events \citep[e.g.][]{Cote2003,Schuberth2010,Ro12,Lo15},
while searches for apparent photometric disturbances such as shells
and tidal tails have also made strong cases for recent and ongoing
accretion \citep{Tal2009}. Extended sheets or pancakes of satellite
galaxies have also been tentatively identified in the outskirts of
clusters~\citep{Fa13}.

Here, we introduce a new method to identify members of the same
infalling satellite galaxy group in a cluster, using only projected
galactocentric distances and line-of-sight velocities. We argue that,
in the outer parts of clusters, galaxies are falling in on almost
radial orbits. This suggests an appealing simple ansatz, namely that
the distribution function (DF) is
\begin{equation}
F \propto \delta(v_\theta)\delta(v_\phi) f(E)
\label{eq:dfradial}
\end{equation}
where $v_r, v_\theta,v_\phi$ are velocity components resolved with
respect to the spherical polar coordinates and $E$ is the energy.  In
Section 2, we show that this leads to an Abel inversion, which can be
performed exactly. In other words, for any tracer density of objects,
a DF of this form can be found, although -- as always -- we must check
for positive definiteness a posteriori.  As an application, Section 3
applies the method to the dataset of satellite galaxies around M87,
one of the most massive galaxies in the local universe.  This giant E0
elliptical resides at the centre of the Virgo cluster, and its
environment has been catalogued extensively by \citet{Binggeli1985}
and \citet{Ki11}. Here, we use the carefully-selected subsample of
Virgo galaxies considered to be certain M87 satellites, compiled by
\citet{Ol15}, and identify possible substructures by looking for
objects which are clustered in energy space, and hence falling onto
M87 on the same orbital path but with different phases. Finally, we
summarise in Section 4 and consider possible extensions, applications
and future prospects for our method.

\section{Radial Orbit Models}
\label{sect:maths}

We consider spherical densities of tracer satellites all moving on
exactly radial orbits.  The DF must depend on the integrals of motion
(Jeans theorem). These are binding energy $E$ and square of the
angular momentum $L^2$ given by
\begin{eqnarray}
E &=& \psi(r) - \frac{1}{2} (v_r^2 + v_T^2)\nonumber\\
L^2 &=& r^2 v_T^2
\end{eqnarray}
where $v_r$ and $v_T$ are the radial and tangential velocity
components judged from the centre of the cluster. The gravitational
(relative) potential $\psi$ is unknown, and we wish to solve for
properties of the potential given the kinematics of the tracers.

Velocity anisotropy is usually characterised by the anisotropy
parameter $\beta = 1 - \langle v_{\rm T}^2\rangle /\langle v_r^2
\rangle$, where angled brackets denote averages over the velocity
distribution. Models with purely radial orbits, such as the ones
presented in this paper, have $\beta =1$ everywhere. While this
assumption is clearly an idealisation, many simulations of
hierarchical formation do produce very strongly radially anisotropic
structures -- for instance, the galaxy haloes formed by \cite{Bu05}
have $\beta \approx 0.8$, as can be clearly seen in Figure 8 of
\citet{Wi15}. Similarly, simulated galaxy clusters have been found to
have strongly radial anisotropy parameters $\beta \approx 0.7$ in the
outer regions, as in Figure 3.2 of~\citet{Sa14}. However, there is
considerable cosmic variance and values $\beta \approx 0.3-0.5$ for
clusters are also present in the literature~\citep[see e.g.,][]{Cu08,
  Wo08, Pr12}. Therefore, the assumption we make here is a
simplification that applies to some, but by no means all, clusters.

\subsection{The general case}

For any tracer density $\nu$, the integral equation for the DF
  $F(E,L^2) = F_E(E)\delta(L^2)$ is of Abel form and can be inverted
  in a similar manner to \citet{Ed16}'s classical work to give
\begin{equation}
F_E(E) = {\sqrt{2} \over \pi^2} {d \over dE} \int_0^E {d\psi \over
(E-\psi)^{1/2}} r^2 \nu, 
\label{eq:abel}
\end{equation}
where $r^2\nu$ is regarded as a function of $\psi$ via the
inversion of $\psi(r)$. This is the general solution for the DF of
  any radial orbit model. Note that the well-known Osipkov-Merritt
  models~\citep[see e.g.,][]{BT} attain $\beta =1$ or extreme radial
  anisotropy at large distances, and so are closely related to our
  models.

The problem is straightforward if the function $F_E$ is a
  power-law, so that the DF has the form
\begin{equation}
F(E,L^2) = \begin{cases}F_0 \delta(L^2) (E-E_0)^{-p} & E \ge
  E_0 \cr 0 & E < E_0,
\end{cases}
\end{equation}
where $F_0$ is a normalisation constant. The model has a finite edge
$r_{\rm t}$ at which $\psi(r_{\rm t}) = E_0$. The density is
obtained by integrating over velocity space as
\begin{equation}
\nu(r) = {\pi^{\threeh} \Gamma(1 - p) 
\over \sqrt{2}\Gamma(\threeh -p)} {F_0\over r^2 (\psi - E_0)^{p\!-\!\half}},
\label{eq:nudens}
\end{equation}
with $p<1$. By integrating along the line of sight, we obtain the
surface density $\Sigma(R)$ as:
\begin{equation}
\Sigma(R) = {\sqrt{2} \pi^{\threeh} \Gamma(1\!-\!p) F_0 \over \Gamma (\threeh\!-\!p) }
\int_R^{r_{\rm t}} {dr \over r(r^2\!-\!R^2)^{\half} (\psi\!-\!E_0)^{p\!-\!\half}}.
\end{equation}
The intrinsic velocity dispersions generated by this model are
\begin{equation}
\langle v_r^2 \rangle  = {2 \over 3-2p}
 \left( \psi -E_0 \right ), \qquad
\langle v_\theta^2 \rangle = \langle v_\phi^2 \rangle = 0.
\end{equation}
There is no dispersion in the angular velocities, as the satellite
galaxies all move on radial orbits. The square of the radial velocity
dispersion is proportional to the gravitational potential.

The data on any satellite comprise projected positions $R$
  and line-of-sight velocities $v_{\rm los}$.  Let ${\cal P}$ be
  shorthand for the parameters of the model.  Using the product rule
  for probabilities, we have:
\begin{equation}
P(v_{\rm los}, R | {\cal P}) = P(v_{\rm los} | R, {\cal P}) P(R |
{\cal P}).
\end{equation}
Here, $P(R | {\cal P})$ is the probability of finding a
  satellite galaxy at projected position $R$, which is just $2 \pi R
  \Sigma(R)$, modulated by the selection function $S(R)$ of
  the survey. The probability distribution of line-of-sight velocity
  $v_{\rm los}$ at any position is
\begin{equation}
P(v_{\rm los}| R, {\cal P}) =  {2\pi F_0\over \Sigma(R)} J(v_{\rm
  los}, R),
\end{equation}
with
\begin{equation}
J(v_{\rm los}, R) = \int^{r_{\rm t}}_{\rmin} {dr \over
  [(r^2\!-\!R^2)(\psi\!-\!E_0)\!-\!\frac{1}{2} v^2_{\rm los} r^2]^p
  (r^2\!-\!R^2)^{1\!-\!p}}.
\end{equation}
Here, $\rmin$ is the smallest positive root for $x$ of
\begin{equation}
  (x^2-R^2)(\psi(x) -E_0) - \frac{1}{2} v^2_{\rm
    los} x^2 =0.
\end{equation}
The physical significance of $\rmin$ is that it is the minimum
three-dimensional position at which we can find a radial orbit whose
velocity projected along the line of sight is $v_{\rm los}$. 

Given a set of $N$ satellite galaxies with projected positions $R_j$
and line of sight velocities $v_{{\rm los},j}$, the logarithm of the
likelihood is then
\begin{equation}
\log L = \sum_{j=1}^{N} \log P(v_{{\rm los},j}, R_j| {\cal P}),
\end{equation}
whose maximum needs to be found via a grid search to identify the best
model parameters ${\cal P}$.  If the spatial selection function $S(R)$
is not known, we can still make progress by taking the likelihood as
\begin{equation}
\log L \approx \sum_{j=1}^{N} \log P(v_{{\rm los},j}| R_j, {\cal P}),
\label{eq:like}
\end{equation}
which is tantamount to assuming that the surface density falls
approximately like $R^{-1}$ and makes only a modest contribution to
the likelihood.  We check the validity of this assumption a
posteriori.

\subsection{Power-law tracers}

So far, we have described the method for an arbitrary spherical
potential. Now, let us consider a specific case which can be applied
to real datasets. We assume that the potential $\psi$ behaves like a
power law
\begin{equation}
\psi = \psi_0 \left( {r_0\over r } \right)^\alpha,
\end{equation}
so that (assuming $r_{\rm t} \rightarrow \infty$ so $E_0
  \rightarrow 0$)
\begin{equation}
\nu(r) = N_0 \left ({r_0\over r} \right)^\gamma,
\end{equation}
where the normalisation constant $N_0$ is given in Appendix A.  We
have put $p = \frac{(2-\gamma)}{\alpha} + \frac{1}{2}$, so that the
absolute value of the logarithmic gradient of the potential is
$\alpha$ and that of the density is $\gamma$.

The observables -- the surface density and the line-of-sight velocity
dispersions -- are also power laws given by
\begin{eqnarray}
\Sigma(R) &=& S_0   \left ({r_0\over R} \right)^{\gamma -1},\nonumber\\
\langle v_{\rm los}^2 \rangle &=& v_0^2 \left ({r_0\over R} \right)^{\alpha},
\label{eq:observables}
\end{eqnarray}
where again the normalisation constants $S_0$ and $v_0$ are relegated
to the Appendix. Our model therefore describes a population of
satellite galaxies whose density and velocity dispersion profiles are
power laws to at least a reasonable approximation. This is reasonable
given that satellite galaxies mainly reside in the outer parts. For
the same reason, we also do not worry about the singularity of the
power-laws at $r=0$.

In this model, the distribution of line-of-sight velocities (the
line profile) is
\begin{equation}
P(v_{\rm los}| R, {\cal P}) = {2\pi F_0\over \Sigma(R)} J(v_{\rm
  los}, R),
\label{eq:lp}
\end{equation}
with
\begin{equation}
J = \int^{r_{\rm
    t}}_{\rmin} {r^{2-\gamma +\alpha/2} dr \over
  [\psi_0 r_0^\alpha(r^2-R^2) - \frac{1}{2} v^2_{\rm los} r^{\alpha+2}]^p (r^2-R^2)^{1-p}}.
\label{eq:lp}
\end{equation}
As before, $\rmin$ is the root of 
\begin{equation}
  (x^2-R^2)\psi_0 r_0^\alpha - \frac{1}{2} v^2_{\rm los} x^{\alpha+2}
  =0.
\end{equation}
This enables us to construct the logarithm of the likelihood according
to Equation~(\ref{eq:like}), and therefore find the best fit of our model
to data.

As an aside, we briefly note that the case $p=\half$ is a singular
limit and eq.~(\ref{eq:nudens}) becomes independent of the
potential. The model is then the exact solution discovered by
\citet{Fr84} and reported in \citet{BT}. Choosing $F_0 = \sqrt{2} C
/\pi^2$, the model then has a density given by
\begin{equation}
\nu(r) = \begin{cases} {\displaystyle C\over \displaystyle r^2} & r <r_{\rm t} \\
                        0 & r> r_{\rm t}.
\end{cases}
\end{equation}
Unlike the models described earlier in this Section, this DF solves
Equation~(\ref{eq:nudens} in a self-consistent way, such that the potential
and density are related via Poisson's equation. As is well-known, such
self-consistent models are subject to the radial orbit instability;
however, we emphasise that this does not apply to our models in which
the distribution function describes a tracer population and so
stability is not compromised.

\begin{figure}
\centering \includegraphics[trim=20 10 10
  20,clip,width=0.5\textwidth]{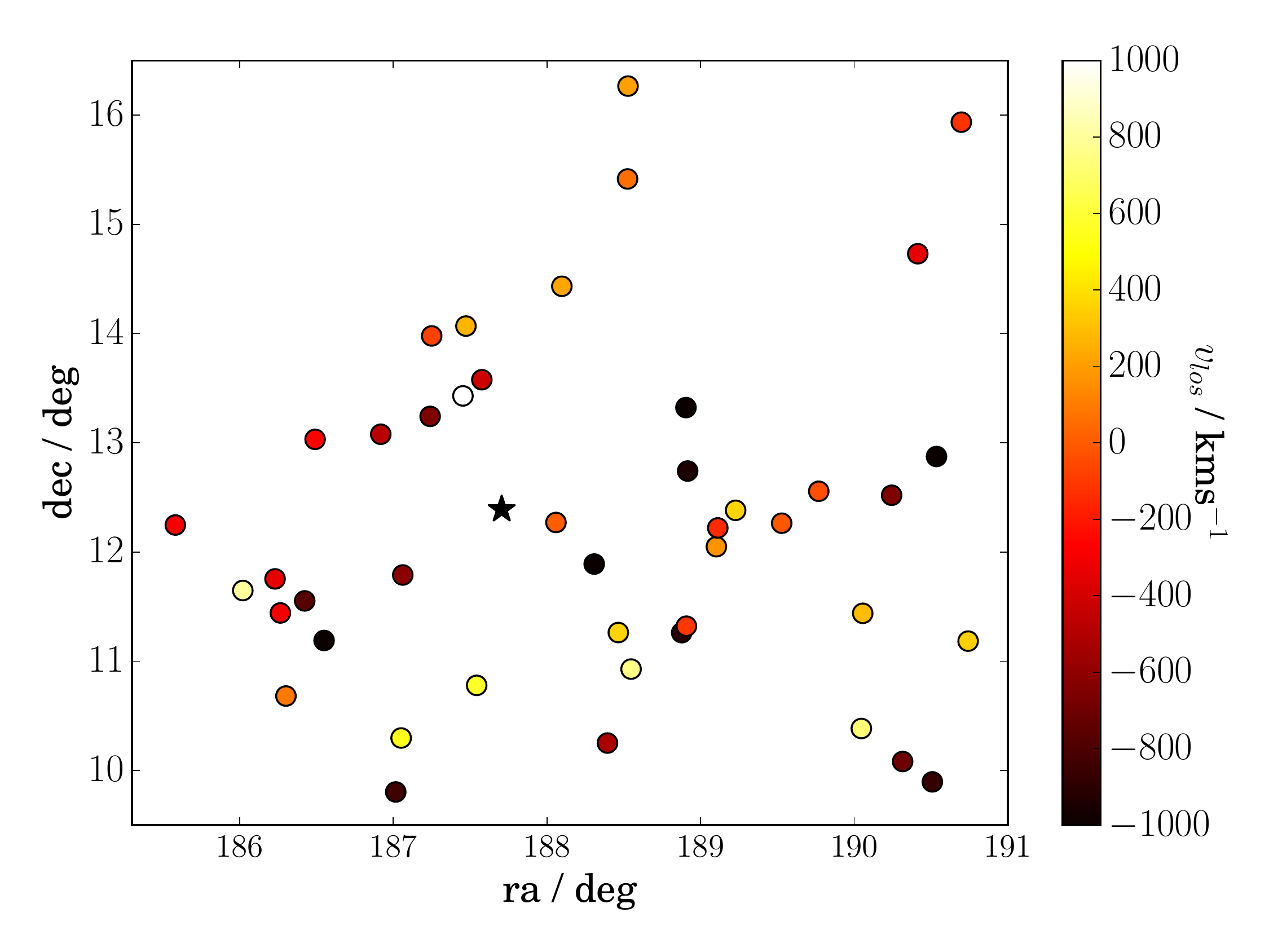}
\caption{A map of the satellite galaxies, colour-coded by
  line-of-sight velocity. M87 is shown by a black star.}
\label{fig:evcc}
\end{figure}

\section{Case Study: M87}

\subsection{Data}

We demonstrate the working of this method by applying it to the
population of satellite galaxies belonging to M87, the BCG in the
Virgo cluster, located at redshift $z = 0.004$ or a distance $D_L =
16.5$ Mpc \citep{Mei2007}. Though Virgo is a relatively poor cluster
compared to others (e.g. Coma), its proximity to us means it is
perhaps the best studied: \cite{Binggeli1985} were the first to
compile an extensive catalogue of possible cluster members, comprising
2096 galaxies within an area of $\sim 140 $ deg$^2$, including
velocities for 572 objects, and classifying 1227 galaxies as certain
cluster members. More recently, \cite{Ki11} presented an Extended
Virgo Cluster Catalogue (EVCC) based on the Sloan Digital Sky Survey
(SDSS) Data Release 7, increasing the area coverage by a factor of
5. The EVCC includes 1589 cluster candidates -- 1324 of which have
velocities measured from the SDSS spectra, with the remainder having
velocity measurements taken from Ned Wright's Extragalactic Database
(NED) -- with each classified as a certain or possible cluster member
according to morphological and spectroscopic criteria. However, Virgo
has a somewhat complex environment comprising a number of sub-clumps,
with two main `clouds' -- the A cloud, centred on M87 and the B cloud,
centred on M49, both at similar line-of-sight distances -- along with
a number of smaller structures, such as the W cloud, which is located
slightly further away at $D_L \sim 23$ Mpc. This means that, to select
a sample of galaxies that are \textit{associated dynamically} with
M87, it is not sufficient to simply take \textit{all} confirmed
cluster members: care must be taken to separate satellites belonging
to the different structures, each of which has its own massive central
object in whose potential the galaxies can be taken to move.

We therefore use the (significantly smaller) subset of satellites
compiled in \cite{Ol15}. While the selection criteria were fully
explained in that paper, we briefly summarise them here: objects
classified as certain cluster members in \cite{Ki11} were
cross-correlated with the distance modulus catalogue of \citet{Bl09}
and removed from the sample if $D_L > 20$ Mpc, to avoid contamination
with the smaller, more distant substructures. While this step has a
dramatic effect on the number of objects in the sample, it is
important because it is impossible to tell from redshift measurements
alone whether an object is at the desired distance or instead has a
peculiar velocity which only makes it appear to be. Objects with
declination angles $< 9$ degrees were further removed to avoid
contamination from the B cloud. Here, we apply a further cut in
projected radius $R<800$ kpc, beyond which the potential from the
other subclumps may be having a non-negligible effect on the
dynamics. This leaves a catalogue of 50 satellites ranging from 248
kpc to 794 kpc from the centre of M87, as shown in
Figure~\ref{fig:evcc}. We note that while the EVCC does not provide
uncertainties on the velocity measurements, it does provide a
comparison between SDSS and NED velocities for the 498 objects having
measurements in both, and indicates a mean difference of $\Delta v =
2.6$ kms$^{-1}$. We therefore assume all objects to have velocity
uncertainties comparable to the typical uncertainty in the SDSS
measurements, which we conservatively take to be 4 \%. We also note
that, since the sample in \citet{Ki11} was selected spectroscopically,
its distribution may not be representative of the underlying
population -- it may, for example, be missing the faintest galaxies --
and that further, as we have applied additional rigorous cuts to
eliminate contaminants, our sample is not complete. Nevertheless, we
are able to make progress under the (not unreasonable assumption) that
our sample is a close representation of the parent population.

\begin{figure}
\centering \subfigure{\includegraphics[trim=20 10 20
    20,clip,width=0.235\textwidth]{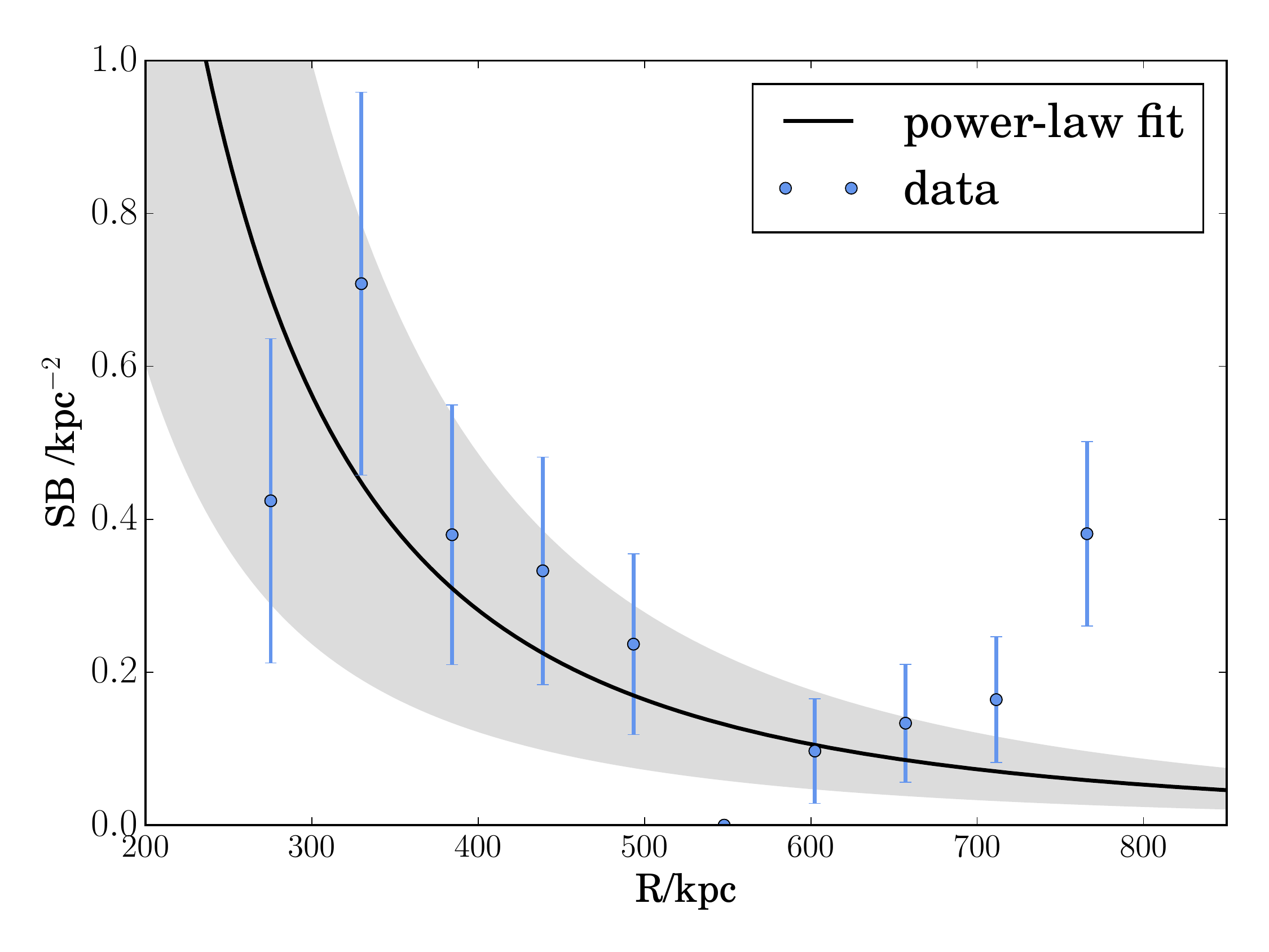}}\hfill
\subfigure{\includegraphics[trim=20 10 20
    20,clip,width=0.235\textwidth]{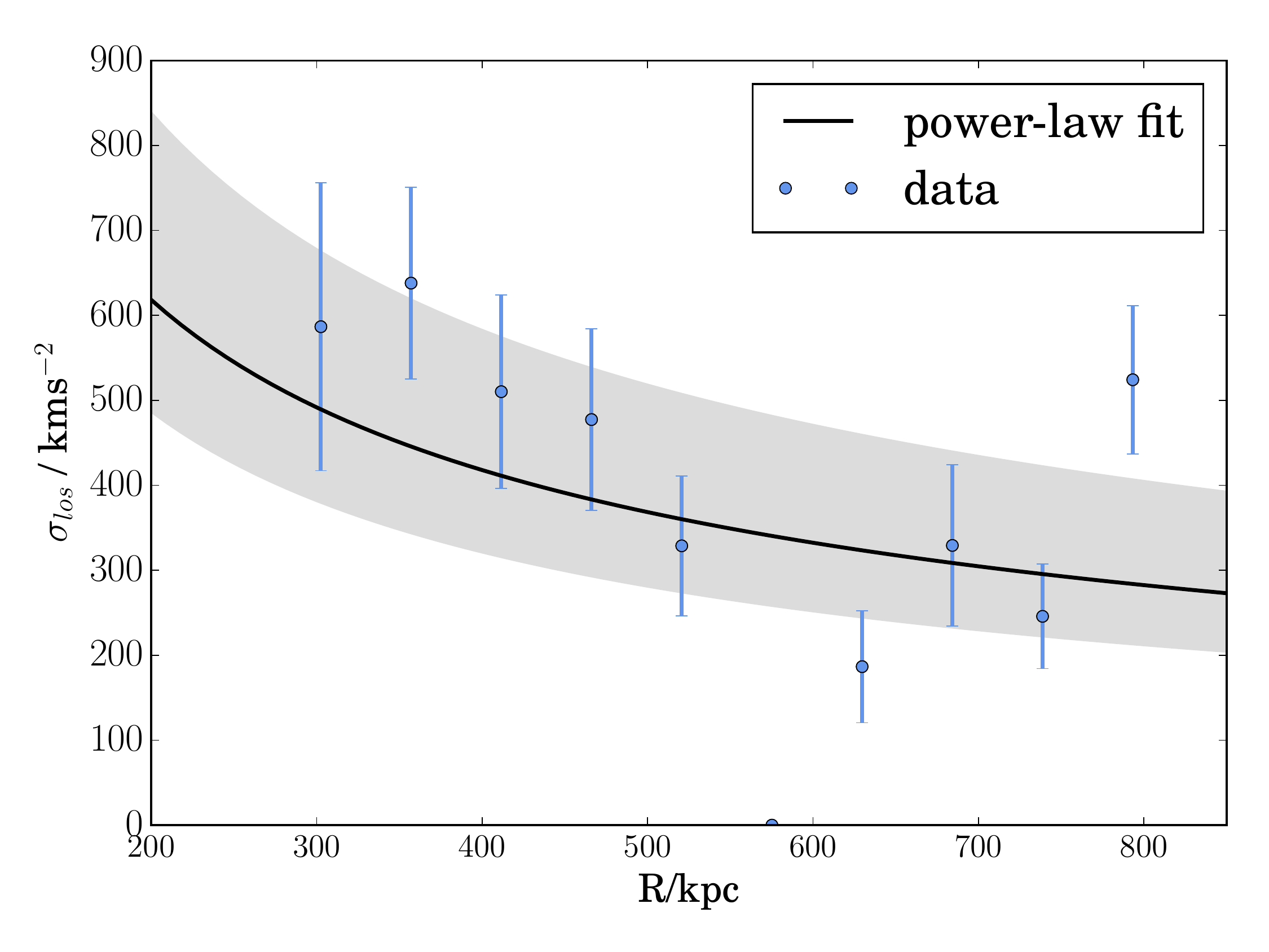}}\hfill
\caption{The binned surface density of satellite galaxies (left) and
  the binned velocity dispersion profile (right), with the fit
  overplotted. The number of objects per bin is small (typically $\sim
  5$). The deviation from the fit at the largest radii may be due to
  small-number statistics, or alternatively may be a sign of the
  limitations of our model.}
\label{fig:vlos}
\end{figure}

\subsection{Implementation: the DF}

To fit to the observed line-of-sight velocities and projected
distances for our sample of Virgo cluster galaxies, we carry out a
grid search for the maximum-likelihood parameters $\psi_0$, $\alpha$
and $\gamma$ of the potential and density profiles, and account for
uncertainties on the parameters due to uncertainties in the data by
Monte-Carlo sampling -- though we note that the projected positions
$R$ have negligible uncertainty. We also allow for the uncertainty due
to the curvature of the likelihood surface. The parameters are
summarised in Table~\ref{tab:fit}, together with 1$\sigma$
uncertainties.  Figure~\ref{fig:vlos} then shows the binned surface
density and line-of-sight velocity dispersion, with our fitted
profiles overplotted. 
This suggests the model is an encouraging match to
the data and validates the assumptions made in our approach. We can
also perform the test of assuming instead a constant selection
function and fully incorporating the spatial information on the
satellite galaxies in the likelihood.  This changes the inference on
both $\alpha$ and $\gamma$ by $\sim 0.1$, which is consistent with our
current uncertainty.

\begin{figure}
\centering
 \includegraphics[trim=20 10 10 20,clip,width=0.5\textwidth]{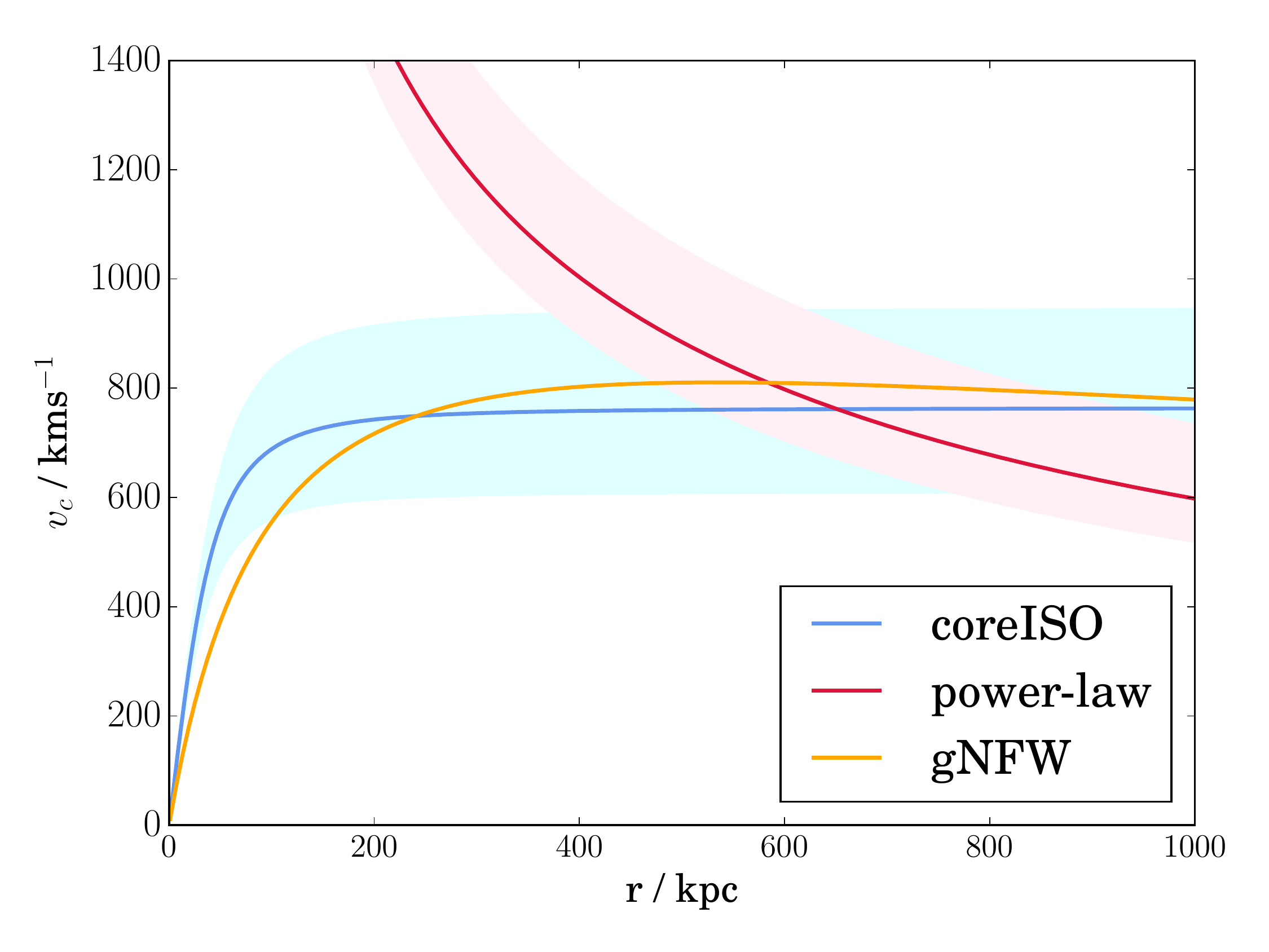}
\caption{A comparison of the circular velocity curves predicted by
  this model and the (cored isothermal and generalised NFW) models of
  \protect\cite{Ol15}. The deviation at small radii is mainly due to
  the fact that the majority of our satellite galaxies are at larger
  radii, meaning we cannot expect to constrain the potential there:
  also, we the power-law potential that we consider here is a simpler
  and more rigid model than in \protect\cite{Ol15}. On the other hand,
  the profiles agree quite well at large radii, with the potential
  inferred in this study slightly lower due to our assumption of
  complete radial anisotropy.}
\label{fig:vcirc}
\end{figure}

To test our inference, we compute the circular velocity profile
predicted by our model and compare it with those inferred in
\cite{Ol15}, in which a more extensive dataset was used (encompassing
stars and globular clusters in addition to the satellite galaxy
population). In that study, models with stars and globular clusters
moving on both isotropic and anisotropic orbits were considered, while
in all cases, the anisotropy of the satellite population was
calibrated using simulations and taken to be $\beta = 0.3$: here,
then, for simplicity, we compare with the isotropic
case. Figure~\ref{fig:vcirc} shows the comparison: due to the
distribution of the satellite galaxies, we expect to be constraining
the potential most strongly at large radii. We see that, while the
large-radius circular velocity profiles are comparable, we ultimately
predict a lower circular velocity than in the isotropic models of
\citet{Ol15}; this can be clearly understood in terms of our very
different assumptions about the anisotropy of the satellites, with the
models considered here assuming totally radial infall as opposed to
the much milder $\beta = 0.3$ used in the \citet{Ol15} analysis.

\begin{table}
\centering
\begin{tabular}{|cccc|}\hline
$\log(\psi_0)$ & $\alpha$ & $\gamma$ & $\log(F_0)$ \\\hline
$8.89 \pm 0.07$ & $1.13 \pm 0.10$ & $2.41 \pm 0.13$ & $-2.59 \pm 0.07$ \\\hline
\end{tabular}
\caption{Maximum-likelihood values for the potential normalisation
  $\psi_0$ and power-law index $\alpha$, the density power-law index
  $\gamma$ and the normalisation of the DF $F_0$, along with
  uncertainties.}
\label{tab:fit}
\end{table}

\subsection{Implementation: clustering}

Given our inferred potential and density parameters, we search for
clustering in energy among the satellite population using the
Kullbeck-Leibler divergence $D_{kl}$ \citep{Kullback1951}. Recently,
this has been used with success by \citet{Sanderson2015} to infer the
gravitational potential of the Milky Way, and more generally is a good
way of quantifying the difference between two distributions or
probability distributions because of its natural interpretation in
terms of probabilistic inference. Also known as the relative entropy,
the $D_{kl}$ from $p(x)$ to $q(x)$ is given by
\begin{equation}
 D_{kl}(p:q) = \int p(x) \log \frac{p(x)}{q(x)} dx,
\end{equation}
which, following \citet{Sanderson2015}, can also be written 
\begin{eqnarray}
D_{kl}(p:q) & = \int p(x)\log p(x) dx - \int p(x)\log q(x) dx \nonumber\\
 & = <\log L(H_p | x)>_p - <\log L(H_q | x)>_p,
\end{eqnarray}
for logarithmic likelihood $\log L$ and hypotheses $H_p$ and
  $H_q$ that the data are drawn from $p$ and $q$ respectively
  (formally, this assumes flat priors). Seen in this light, the
  $D_{kl}$ can be simply understood as the amount of information lost
  in describing data $x$ as drawn from a distribution $q$ instead of
  the true distribution $p$. This allows us to choose contours of
$D_{kl}$ to set confidence intervals: for a Gaussian probability
distribution, the $1 \sigma$, $2 \sigma$ and $3 \sigma$ levels
correspond to $D_{kl} = 0.5, 2$ and $4.5$ respectively.

For each object, we use the inferred potential, along with the
observed projected position $R$ and line-of-sight velocity $v_{\rm
  los}$, to compute the energy profile as a function of the unknown 3D
radius $r$, as
\begin{equation}
 E(r) = \frac{1}{2}\frac{r^2 v_{\rm los}^2}{r^2-R^2} - \psi(r),
\end{equation}
where we have used the fact that all motion is radial, such that $r^2
v_{\rm los}^2 = (r^2-R^2) v_r^2$ . This can be inverted to give the
position $r = r(E)$ as a function of radius, and the probability
$p(E|R_p,v_{\rm los},{ \cal P})$ of finding an object with energy $E$
given its observed $R$ and $v_{\rm los}$ and model parameters ${\cal
  P} = (\psi_0,\alpha,\gamma)$ can be calculated as
\begin{equation}
 p(E|R,v_{\rm los},{\cal P}) \propto 4 \pi r^2 \nu(r) \Bigl| \frac{dr}{dE} \Bigr|.
\end{equation}
We note that the probabilities are very peaked due to the stationary
points in the energy distributions, which allow the energy to be
minimised at some $r$. This preference for the minimum-energy
configuration also results in the probability distributions being
asymmetric, with the maximum probability always associated with the
smallest allowable energy. This suggests that the probability
distributions themselves could be well-approximated by truncated
Gaussians, where the truncation is applied at the peak of the
distribution itself. Figure~\ref{fig:clumps} shows the energy
distributions under this approximation, for the subset of objects
which are found to belong in groups.

We therefore model the probability distributions using truncated
Gaussians with full-width half-maxima (FWHM) determined by the
energies at which the true distributions drop to half their peak
value, and compute the $D_{kl}$ analytically: we check that this is a
good approximation by also fitting the distributions using cubic
splines and integrating the $D_{kl}$ numerically, and find that the
effect of this is negligible. We also test an even simpler model in
which the distributions are described by (non-truncated) Gaussians,
though here we find that the loss of asymmetry in the energy
distributions causes significantly more clumping to be found: this
indicates the importance of modelling the energy distributions using
appropriate parametric forms. While our sample here is small, it is
important to check that fast analytic representations of the
probability distributions are sufficient, as this means that the
method can be tractably applied to much larger datasets without the
need for numerical integration. For each object, we then convolve the
FWHM from the probability distribution with an uncertainty due to our
assumed measurement uncertainty of 4\% on $v_{\rm los}$, in addition
to a correction to allow for the fact that the objects may not be
moving completely radially: to this end, we broaden the distribution
to account for an uncertainty $\Delta \beta = 0.2$ on the
anisotropy. We also investigate the dependence of the clump
identification of the upper cut-off in $D_{kl}$ by comparing the
clustering that is identified at both the $1 \sigma$ and $2 \sigma$
levels.

\begin{figure}
\centering
\includegraphics[trim=20 20 20 20,clip,width=0.5\textwidth]{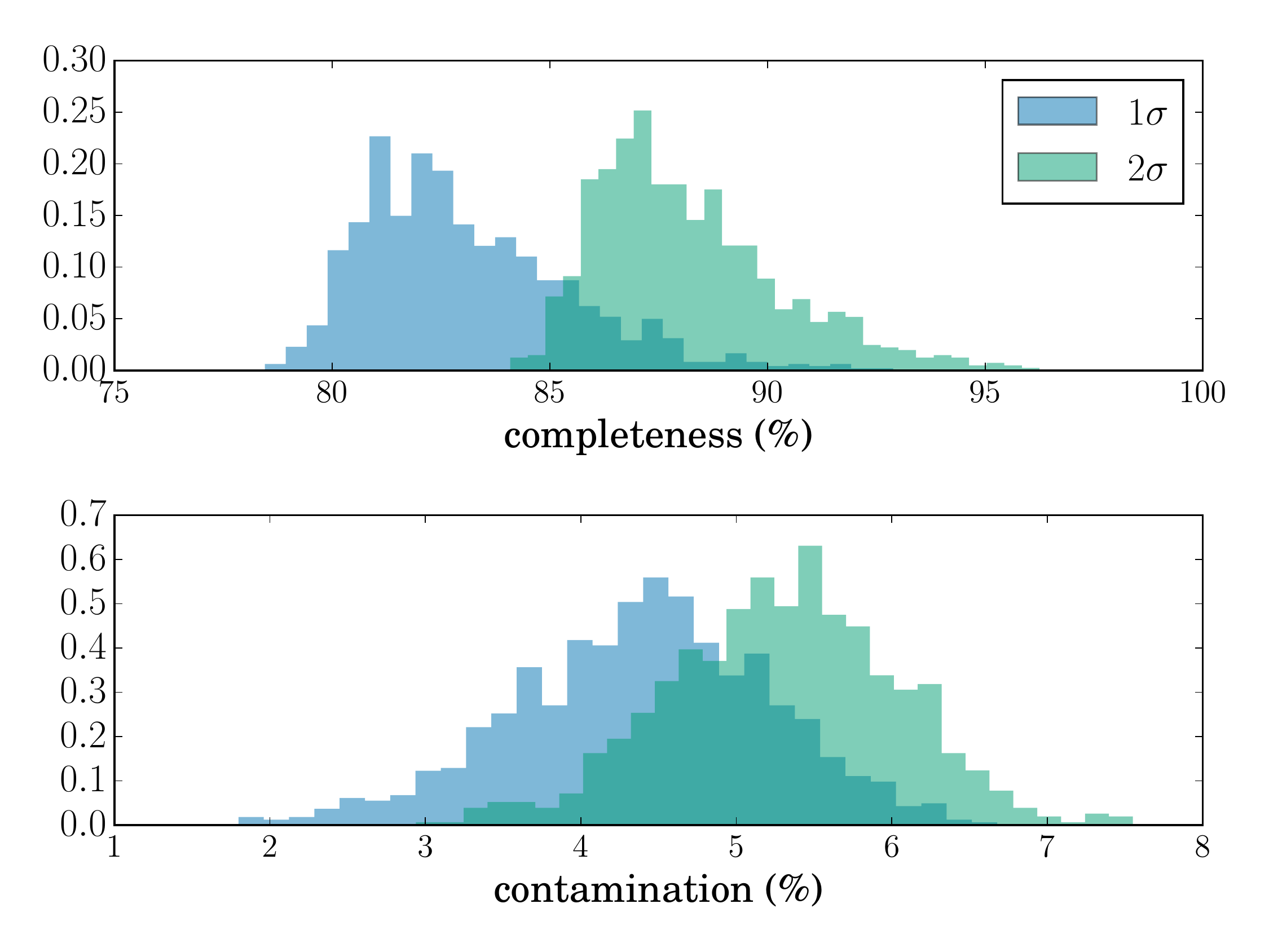}\hfill
\caption{Mock catalogues generated from the DF, with single groups of
  between three and seven objects inserted, are analysed using our
  $D_{kl}$ algorithm and, as shown by these normalised histograms,
  imply high levels of completeness and low levels of contamination
  for both $1 \sigma$ and $2 \sigma$ cut-off levels. Given that the
  contamination fraction changes little between the two limits,
  whereas the more generous limit allows a significantly higher
  completeness, we adopt the $2 \sigma$ limit for the majority of this
  work.}
\label{fig:histograms}
\end{figure}

Initially, we test the ability of the $D_{kl}$ analysis to identify
`real' structure. To do this, we generate 1000 mock datasets, each the
same size as the real dataset, using random draws from the DF, and
insert single groups of between three and seven objects which have
themselves been drawn from a Gaussian PDF with some central energy and
uncertainty, both of which are chosen to be within the 20th and 80th
percentiles of those of the main mock dataset. We then perform our
$D_{kl}$ test on each mock catalogue. The result is summarised in
Figure~\ref{fig:histograms}: for cut-offs in $D_{kl}$ at both the $1
\sigma$ and $2 \sigma$ levels, we find that the contamination fraction
is consistently low, with an average of $\sim 5 \%$; for the $1
\sigma$ test, the average completeness is $82 \%$ while for the $2
\sigma$ test, this increases to $87 \%$. In both cases, the scatter is
small, at $\sim 9 \%$. This is reassuring, as it indicates that our
$D_{kl}$ analysis is indeed a good way of picking out clustering in
this scenario; further, given the result that the $1 \sigma$ and $2
\sigma$ tests both have similar levels of low contamination, we
investigate both cut-offs in what follows but ultimately pursue only
the $2 \sigma$ results.

\begin{figure*}
\centering
 \subfigure{\includegraphics[trim=50 20 25 20,clip,width=0.5\textwidth]{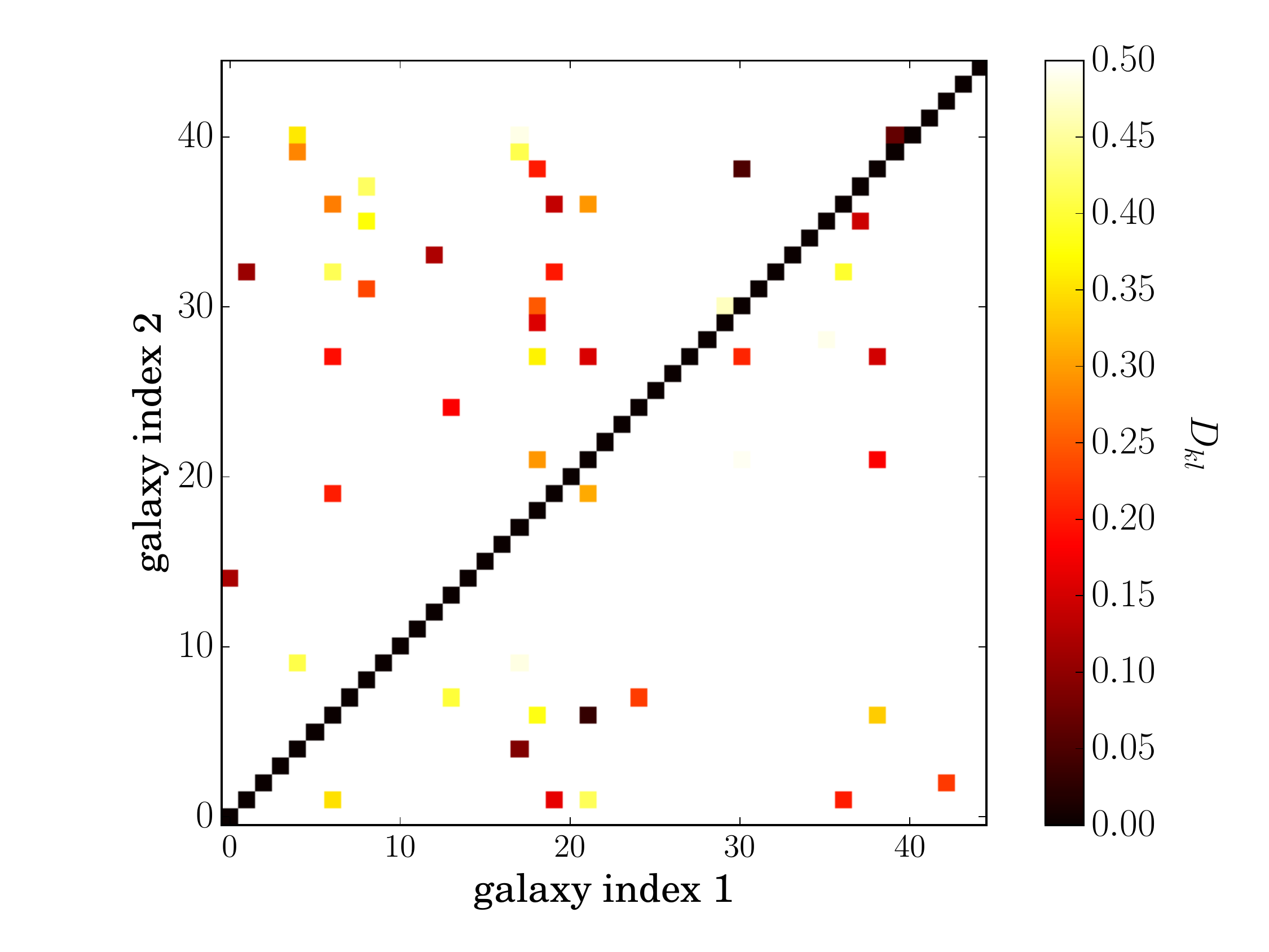}}\hfill
 \subfigure{\includegraphics[trim=50 20 30 20,clip,width=0.5\textwidth]{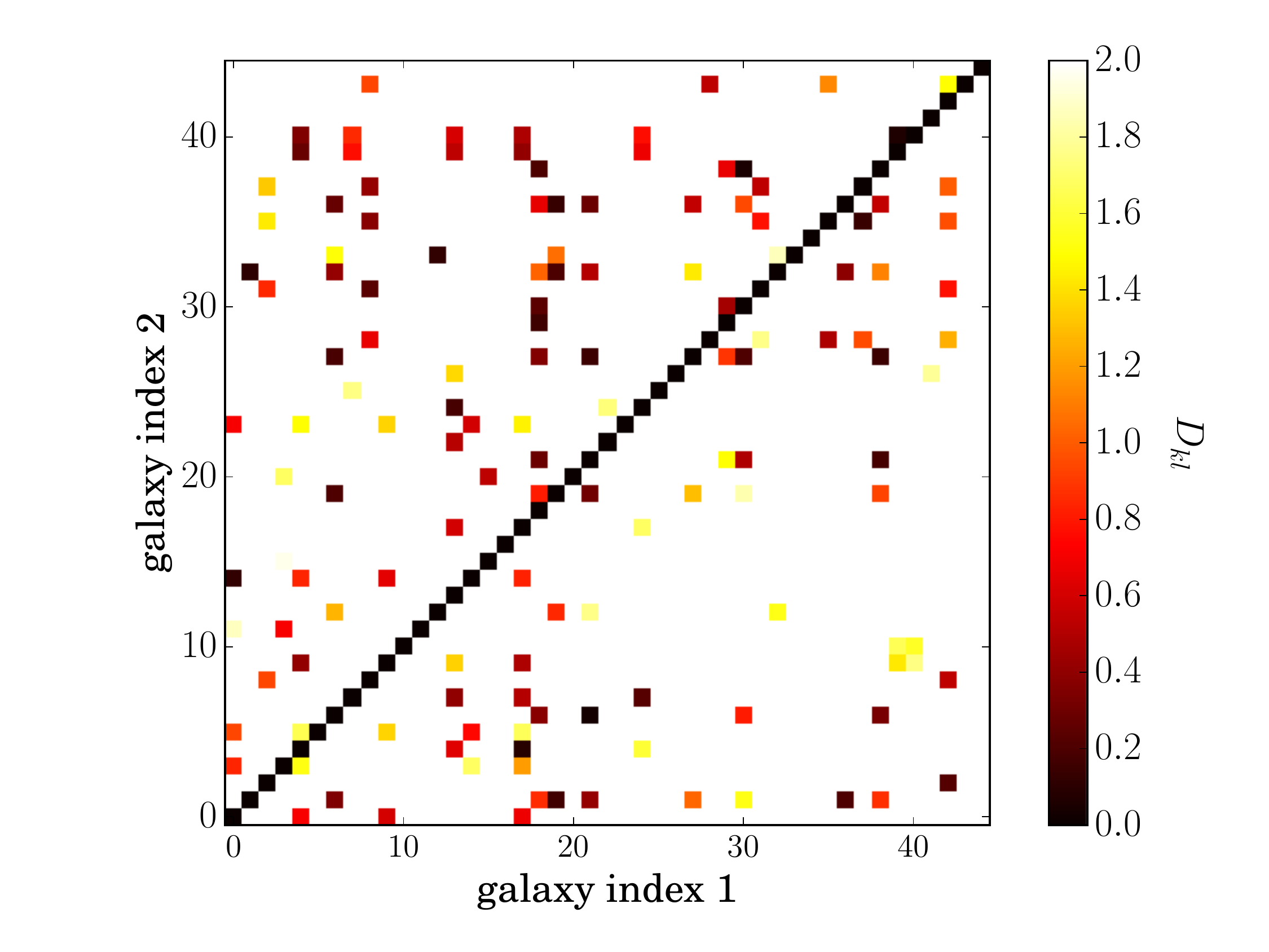}}\hfill
\caption{The Kullback-Leibler divergence $D_{kl}$ between all galaxy
  pairs. For identical distributions, $D_{kl} = 0$, as can be seen
  down the diagonal. For an overlap between the two distributions $< 1
  \sigma$ for a Gaussian distribution, $D_{kl} < 0.5$. Hence coloured
  squares are indicative of pairs that may have been drawn from the
  same initial energy distribution. Left: with a $1\sigma$ cut-off in
  $D_{kl}$. Right: with a $2\sigma$ cut-off.}
\label{fig:tshirt}
\end{figure*}

\begin{figure*}
\centering
\subfigure{\includegraphics[trim=20 10 10 20,clip,width=0.5\textwidth]{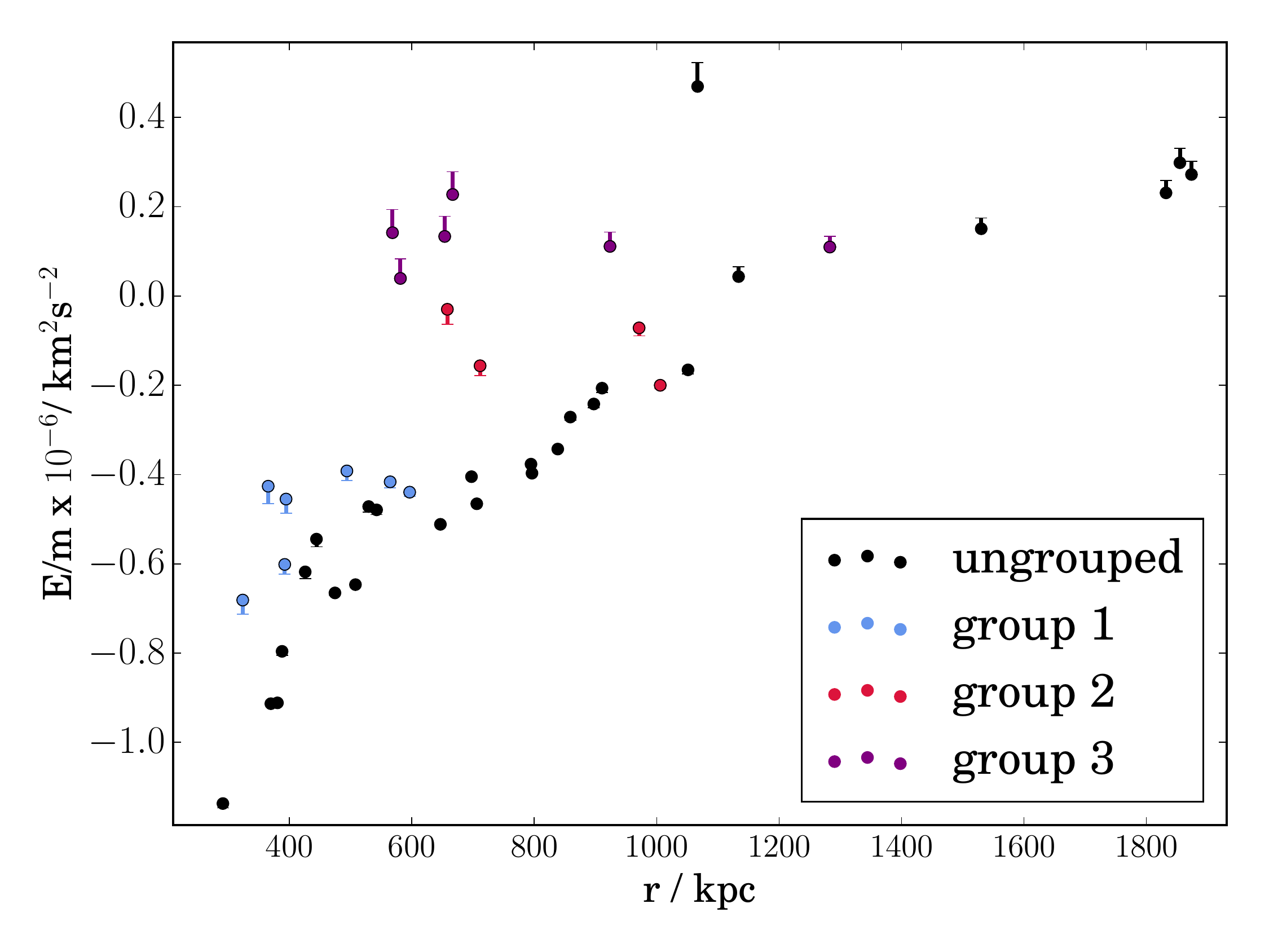}}\hfill
\subfigure{\includegraphics[trim=20 10 10 20,clip,width=0.5\textwidth]{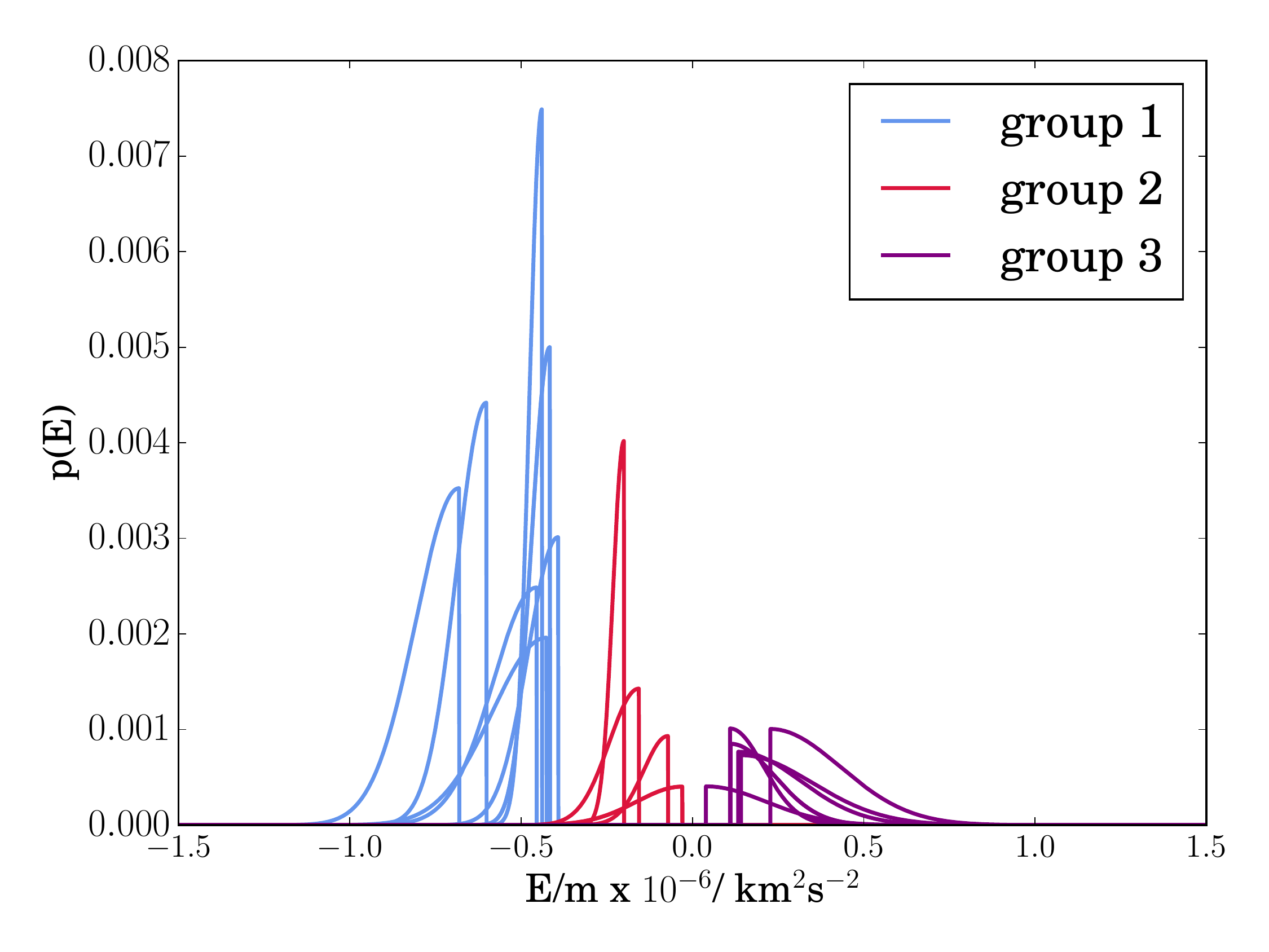}}\hfill
\caption{Left: Energy-position maps of the satellites, colour-coded
  according to the identified groups for the $2 \sigma$
  case. For each satellite, we use the energy and radius
    corresponding to the maximum of the probability distribution.
  Right: Probability distributions in energy for the satellites,
  colour-coded in the same groups. The clumps are much easier to pick
  out in this 1D space.}
\label{fig:clumps}
\end{figure*}

Figure~\ref{fig:tshirt} presents the $D_{kl}$ grid for all the
different galaxy pairs for both $1 \sigma$ and $2 \sigma$ limits,
labelling each galaxy by its index in our catalogue. We note
  that the asymmetry of the energy distributions and in the definition
  of the $D_{kl}$ carry through to an asymmetry in divergence space,
and that $D_{kl} = 0$ along the diagonal as there is no information
lost in describing the energy distribution of galaxy $i$ as being
drawn from the energy distribution of galaxy $i$. It is also clear
that a larger degree of clustering is identified when the divergence
limit is relaxed from $1 \sigma$ to $2 \sigma$, as this is a more
generous criterion for associating energy distributions --
  this is consistent with our finding that the $2\sigma$ limit allows
  a higher level of completeness without significantly compromising
  the level of contamination.  Where a set of objects emerges
  with mutually small divergences, these can be sorted into a `group'
  or cluster; however, we note that the $D_{kl}$ strictly only sets
  upper limits on sets of groupable objects, as opposed to identifying
  groups. That is, given two objects with a $D_{kl} < 2 \sigma$, we
  can formally only say that these are compatible with being in the
  same group at the 2$\sigma$ level, given current observational
  limitations.

Table 2 summarises the groups that are identified in the $2
  \sigma$ case, and their distributions in both the 2D energy-position
  space and 1D energy space are shown in the middle and lower
  left-hand panels of Figure~\ref{fig:clumps}: for the former, we note
  that objects that are clustered in energy can be very dispersed in
  their positions on the sky, as might be expected to arise from phase
  mixing over the time since they started their infall, and that
  clustering can therefore be seen particularly clearly when plotted
  as a function of energy alone. With this $2 \sigma$ limit, we find
  three clearly distinct clusters of galaxies with energies that
  appear to have been drawn from the same distribution, each with a
  small number of members (4, 6 and 7 objects); a remaining 28 objects
  do not appear to be associated with any others in this paradigm. We
  note that when we restrict our divergence limit to $1 \sigma$, we
  find that the original groups fragment, and we end up with five
  clusters, each containing either two or three objects. 

As a final test, we also examine the uncertainty introduced
  into our $D_{kl}$ analysis via uncertainties in $\alpha$, $\gamma$
  and $\psi_0$. We do this by repeating the analysis for 100 samples
  of $(\alpha, \gamma, \psi_0)$ drawn from their uncertainty
  distributions. We find the clustering to be robust against changes
  of the order of the uncertainties in all three parameters, but as
  long as correlations in the PDF between $\alpha$ and $\psi_0$ are
  accounted for correctly. That is, these two parameters are
  degenerate as they combine to describe the potential. Thus, drawing
  samples from the PDF assuming no covariance causes both parameters
  to be sampled too widely and thus begins to introduce uncertainty
  into the final object groupings. However, as long as the covariance
  is accounted for correctly, the same groupings are generally
  recovered at the $2 \sigma$ level (with a small number of objects
  passing in and out of the larger groups across the samples). This is
  reassuring and suggests that our identification of substructure
  should not be significantly affected by our uncertainties on the
  potential and surface density.

\begin{table*}
\centering
\begin{tabular}{cccccccccc}\hline
 ID & ra (deg) & dec (deg) & R (kpc) & $v_{\rm los}$ (kms$^{-1}$) & $u$ (mag) & $g$ (mag) & $r$ (mag) & $i$ (mag) & $z$ (mag) \\\hline
\multicolumn{10}{c}{ Group 1 }\\\hline
1 & 186.92 & 13.08 & 309.76 & -482.40 & 13.01 & 11.29 & 10.48 & 10.10 & 9.91 \\
4 & 187.06 & 11.79 & 248.02 & -609.60 & 15.48 & 13.66 & 12.89 & 12.39 & 12.40 \\
5 & 187.24 & 13.24 & 287.81 & -648.60 & 14.35 & 12.66 & 11.94 & 11.51 & 11.28 \\
14 & 188.39 & 11.26 & 370.03 & -521.40 & 16.61 & 15.56 & 14.85 & 14.49 & 14.25 \\
18 & 188.55 & 12.05 & 257.81 & 750.20 & 16.29 & 14.92 & 14.21 & 13.81 & 13.60 \\
8 & 187.47 & 14.07 & 504.22 & 269.80 & 13.85 & 12.15 & 11.39 & 11.00 & 10.76 \\
25 & 189.23 & 12.52 & 453.82 & 363.85 & 15.91 & 14.25 & 13.63 & 13.33 & 13.05 \\\hline
\multicolumn{9}{c}{ Group 2 }\\\hline
3 & 187.05 & 10.30 & 629.11 & 544.55 & 16.18 & 14.88 & 14.25 & 13.78 & 13.47 \\
9 & 187.54 & 10.78 & 482.31 & 577.25 & 14.86 & 13.43 & 12.77 & 12.40 & 12.31 \\
38 & 185.58 & 14.76 & 784.52 & -299.05 & 16.59 & 15.42 & 14.71 & 14.34 & 14.01 \\
43 & 186.42 & 12.81 & 402.81 & -772.90 & 14.37 & 12.65 & 11.90 & 11.49 & 11.24 \\\hline
\multicolumn{9}{c}{ Group 3 }\\\hline
7 & 187.45 & 13.43 & 318.31 & 1008.35 & 12.63 & 10.78 & 10.00 & 9.59 & 9.32 \\
19 & 188.88 & 12.22 & 342.04 & -902.70 & 13.78 & 12.17 & 11.42 & 11.03 & 10.77 \\
20 & 188.91 & 12.38 & 352.58 & -1023.75 & 16.82 & 15.11 & 14.34 & 13.94 & 13.65 \\
22 & 188.92 & 12.56 & 363.50 & -943.80 & 12.25 & 10.42 & 9.63 & 9.21 & 8.92 \\
31 & 190.32 & 11.39 & 677.70 & -708.75 & 15.92 & 14.17 & 13.40 & 13.01 & 12.78 \\
39 & 186.02 & 11.22 & 505.59 & 800.20 & 14.77 & 13.15 & 12.44 & 12.04 & 11.81 \\\hline
\end{tabular}
\caption{Galaxy clumps identified at the 2$\sigma$ level. Radii are
  clustocentric (taking M87 as the centre of the cluster) and
  line-of-sight velocities are also measured relative to M87;
  magnitudes are taken from SDSS DR7.}
\end{table*}

\section{Discussion and Conclusions}

Strong tests of the current cosmological paradigm are provided by the
abundance of substructure. As larger structures are assembled
hierarchically from mergers and accretion, we should be able to
identify fossil signatures of these events in galaxies and clusters.
In particular, galaxy clusters are characterised by a virialised
region within which all components -- galaxies and dark matter -- are
in dynamical equilibrium surrounded by infall zones in which groups of
galaxies are falling into the relaxed cluster.

The identification of substructure has mainly been studied in the
context of the Milky Way halo. Here, the existence of (in the best
cases) six dimensional phase space coordinates makes the problem
easier.  For example, the use of the actions and
frequencies~\citep{Mc08,Sm09} has been advocated to identify past
merger events in the Milky Way. The problem of the identification of
substructure in projected datasets -- in which only positions on the
sky, line-of-sight velocities and heliocentric distances are available
-- is harder. It is also of much greater interest and applicability,
both to nearby galaxies and to galaxy clusters.

Here, we have introduced a new formalism to describe the dynamical
state of the outer parts of galaxies and galaxy clusters. The
infalling motions of objects are assumed to be generated by purely
radial orbits.  This means that the probability distributions of
observable quantities can be inferred, given the density of the
infalling tracers and an estimate of the gravitational potential in
which they move. We have provided a general algorithm to do this for
spherical potentials.  This enables us to search for infalling groups
of objects, even though they may be scattered across the sky.
Algorithms to quantify substructure in projected data are scarce. The
only other one known to us looks for shells, which can be quantified
by the characteristic ``chevron pattern'' discernible in line-of-sight
velocity and position plots~\citep{Ro12}.

As a practical application of our method, we have examined the dataset
of satellite galaxies around M87. The extended envelope of M87 has
been built from a deluge of smaller satellite galaxies, which may have
accreted along preferred directions. Hence, we expect correlations in
the satellite galaxy dataset, as some of the satellites may have
fallen in along the same orbital path. Our algorithm exploits these
correlations to identify kinematically similar substructure. In the
case of M87, we have identified a number of possible galaxy
associations. These are satellite galaxies whose position and
kinematics are consistent with infall on the same radial path.  This
is expected in theories like $\Lambda$CDM in which the infall of
satellites is coherent rather than random. This provides proof of
principle that our algorithm can be applied to real data to extract
useful results.

A possible test of the galaxy associations around M87 may be afforded
by deep photometry of the candidates to find the position angles of
the major axis. Infalling objects are expected to be stretched out
along the orbital path, and so -- if they lie on the orbits
conjectured in this paper -- they will be radially distended and their
major axes will point towards M87. This effect is seen in the
accretion of subhaloes in dissipationless simulations~\citep{Ku07,
  Ba15} and persists with the inroduction of
baryons~\citep{Kn10}. Radial alignment has also been detected
observationally in galaxy clusters and groups~\citep[see
  e.g.,][]{Ha75, Ag06}, though the magnitude of this effect is
unclear.  However, we would predict significant isophotal alignment of
the major axis of our candidates with the cluster radial direction if
our associations are real.

Another natural application is to globular cluster and satellite
galaxy datasets in other nearby galaxies. There are intense
observational efforts focussing on completing the surveys of stellar
streams and substructure around the Milky Way; however, a complete
picture can only be obtained by studying a wider sample of galaxies at
greater distances, although this is a much harder problem
observationally.  A good place to start would be with the satellites
and clusters of M31, where coherent streams are readily visible;
slightly further afield, the Centaurus group may also be a good
candidate. We also note that M87 has a very large number of
  globular clusters \citep{Ol15a} which may provide additional
  insights into the accretion history of the galaxy we have studied in
  this work. We anticipate that this algorithm will be a valuable
tool in helping to investigate the build up of structure in the Local
Group and beyond.

It is surprising that DFs built from only radial orbits have not
received much more attention. Perhaps this is because for the fully
self-consistent problem (in which the density generates the
potential), such DFs fall foul of the radial orbit
instability~\citep{Fr84}. However, this objection does not apply to
tracer populations, which are moving in an external potential provided
largely by other stellar and dark matter populations. We have shown
that the radial orbit DF can always be found by Abel transforms via an
inversion similar to \citet{Ed16}'s classical work for the isotropic
DF. In fact, radial orbit DFs are applicable to a wide range of
astrophysical problems. In this paper, we have concentrated on
material infalling onto BCGs, but the DFs are also applicable to
populations expelled from central nuclei.  The hypervelocity stars in
the Milky Way are believed to be ejected by the central black hole
with speeds from a few hundred to a few thousand kms$^{-1}$. The
runaway stars are formed when one component of binary receives a kick
as its companion explodes as a supernovae. Both hypervelocity and
runaway stars are ejected from the central parts of the Milky Way with
such high velocities that they move on almost radial orbits, as shown
by simulations of their space motion by ~\citet{Ke14}. Our DFs should
have a ready application to the descriptions of these radially ejected
populations as well.

\section*{Acknowledgments}
We thank the anonymous referee for their helpful and insightful
comments.  LJO also acknowledges financial support from the Science
and Technology Facilities Council of the United Kingdon.

\appendix

\section{Numerical Constants}

This appendix gives the three normalisation constants for power-law
tracers in power-law potentials moving on radial orbits (see Section 2.2)
\begin{eqnarray}
N_0 &=& 
{\pi^{3/2} \Gamma( [2\gamma + \alpha -4]/(2\alpha) ) \over
\sqrt{2} r_0^2 \Gamma([\gamma + \alpha-2]/\alpha)}
F_0\psi_0^{(\gamma-2)/\alpha} \nonumber \\
S_0 &=& {\pi^2 \Gamma( [\gamma -1]/2 ) \Gamma( [2\gamma+\alpha-4]/(2\alpha) )
  \over \sqrt{2}r_0 \Gamma(\gamma/2)\Gamma([\gamma + \alpha -2]/
    \alpha)}F_0\psi_0^{(\gamma-2)/\alpha} \nonumber \\
v_0^2 &=& {\psi_0 \Gamma(\gamma/2)\Gamma([\alpha + \gamma
    -1)/2] \over 2\Gamma((\gamma-1)/2)\Gamma([\alpha +
    \gamma+2]/2)}.
\end{eqnarray}

\end{document}